\documentclass{SCIS2025}

\newcommand{\Rmnum}[1]{\expandafter\@slowromancap\romannumeral #1@}

\newcommand{\norm}[1]{\left\lVert#1\right\rVert}
\newcommand{\ket}[1]{\left|#1 \right\rangle}

\newcommand{\ketbra}[2]{\left| #1 \rangle \langle #2 \right|}
\newcommand{\polylog}[1]{\mathrm{polylog}\left( #1 \right)}
\usepackage{threeparttable}

\begin{document}
\ArticleType{RESEARCH PAPER}
\Year{2025}
\Month{}
\Vol{68}
\No{}
\DOI{}
\ArtNo{000000}
\ReceiveDate{}
\ReviseDate{}
\AcceptDate{}
\OnlineDate{}
\AuthorMark{}
\AuthorCitation{}

\title{Quantum-assisted anomaly detection with multivariate Gaussian distribution}{Quantum-assisted anomaly detection  with multivariate Gaussian distribution}

\author[1,2]{Chao-Hua YU}{}
\author[1,2]{Hong-Miao RAO}{}
\author[1,2]{Ying-Pei  WU}{}
\author[1,2]{De-Xi LIU}{}
\author[1,2]{\\Xi-Ping LIU}{}
\author[3]{Lin-Chun WAN}{linchunwan@outlook.com}


\address[1]{School of Computing and Artificial Intelligence, Jiangxi University of Finance and Economics, \\
Nanchang 330032, China}
\address[2]{Jiangxi Provincial Key Laboratory of Multimedia Intelligent Processing, Jiangxi University of Finance and Economics, \\
Nanchang 330032, China}
\address[3]{School of Computer and Information Science, Southwest University, Chongqing 400715, China}


\abstract{ Anomaly detection with multivariate Gaussian distribution, which we refer to as Gassian anomaly detection (GAD), is a prominent task in data mining and machine learning. The core task of GAD is to obtain the mean value vector and the covariance matrix that characterize the probability density function of an unknown multivariate Gaussian distribution used to detect anomalies, which could be time-consuming when addressing a large dataset. Recently, several quantum algorithms have been proposed for GAD with substantial speedup over the classical GAD. However, they all require quantum phase estimation as key subroutines so that their quantum ciruits have long depth and are unfavorable in the noisy intermediate-scale and early fault-tolerant quantum eras. In this paper, we propose a quantum algorithm for GAD biult on arithmetic-free black-box quantum state preparation (AFQSP), which significantly shortens the quantum circuit depth and reduces the burden of quantum hardware. Specifically, we take advantage of AFQSP to estimate the magnitude of every mean value and that of every covariance matrix element in the classical form, and develop a technique referred to as Hadamard sign test to further reveal their signs, so that anomaly detection of any data point can be done immediately on a classical computer at little cost. It is shown that our quantum algorithm for GAD achieves exponential speedup over the classical GAD when handling low-dimensional datasets with well-conditioned data matrices, and is also time competitive compared to the prior quantum algorithms for GAD. Moreover, our algorithm releases the requirements of input data being quantum, mean centered, or feature correlated in the prior quantum algorithms for GAD, meaning that our algorithm is more practical on input data. Our work highlights the role of AFQSP in bringing quantum machine learning closer to reality.}


\maketitle

\section{Introduction}
Quantum computing takes advantages of quantum mechanical principles, such as quantum superposition and quantum entanglement, to accomplish computational tasks. It promises to offer significant computational speedup over classical computing in solving certain problems, such as large integer factoring\cite{Shor94}, unstructured search\cite{Grover97}, and solving linear systems of equations\cite{HHL09}. In the past two decades, especially since the pioneering quantum algorithm for linear equations was proposed by Harrow, Hassidim, and Lloyd (HHL)\cite{HHL09}, quantum computing has been combined with data mining and machine learning, giving rise to an interdisciplinary field referred to as quantum machine learning (QML)\cite{QML17,DWGetal25}. Since its emergence, QML has flourished as a rapidly evolving research domain attracting world-wide attention in both academia and industry. While the research of QML generally includes two aspects: developing quantum algorithms to enhance or accelerate implementation of classical machine learning tasks and classical machine learning algorithms to solve quantum problems(such as quantum many-body problems \cite{CK24}), the majority of QML literature lies in the former. Up to now, a variety of quantum algorithms have been reported to implement various machine learning tasks with substantial computational speedup over their classical algorithms, including data classification\cite{LMR13, PL13, QSVM14, CD16, SFP17, SP18, SK19, HCTetal19, SLWQWG24}, clustering\cite{ABG13,KLLP19,AP21,WSZ22}, linear regression\cite{WBL12,SSP16,Wang17,LZ17,YGLetal19,YGW21,CMP23}, association rule mining\cite{QARM16,Y22}, and dimensionality reduction\cite{QPCA14,YGLW19,DYXL19,SDPC21,FGLX24}, etc.

Anomaly detection (AD) is also an essential machine learning task well studied in QML, which aims to identify data instances or events that deviate from expected behavioral patterns \cite{CBK09}. It finds broad applications in domains such as financial fraud detection\cite{HGY22}, network intrusion detection\cite{ASW21}, and healthcare\cite{FGD21}. Among existing methods for AD, Gaussian-based anomaly detection (GAD)\cite{DM16} stands out as a promising one. Given a training set of $M$ data points $\bm{x}_0,\bm{x}_1,\cdots,\bm{x}_{M-1}\in \mathbb{R}^D$, GAD assumes each data point is in an unknown multivariate Gaussian distribution whose probability density function $p(\bm{x})$ is characterized by the mean value vector $\mathbf{\mu}$ and the covariance matrix $C$, according to which any point $\bm{x}$ with probability density function value $p(\bm{x})$ lower than a threshold is classified as an anomaly. Thus, the core of GAD involves computing $\bm{\mu}$ and $C$ of the training dataset to determine $p(\bm{x})$. However, implementing such a task on a classical computer generally takes time complexity $O(D^2(M+D))$, which could be quite time-consuming when dealing with a large dataset. Therefore, exploring how to apply quantum computing to computationally accelerating GAD is of great importance.

In 2018, Liu et al.\cite{LR18} pioneered the concept of quantum anomaly detection, and proposed two quantum algorithms for AD, namely quantum kernel principal component analysis and quantum one-class support vector machine, both achieving exponential speedups over classical counterparts with respect to the number of features $D$.  To date, a series of quantum algorithms have been proposed for AD, including more recently proposed quantum algorithms for sequence AD \cite{GLP23}, local outlier factor-based AD \cite{GPL23} and angle-based AD \cite{RYW25},  demonstrating varying degrees of quantum computational advantages; see \cite{CMD24} for an overview of quantum algorithms for AD. In light of this, exploring quantum algorithms for GAD has attracted considerable attention. Hereafter, we refer to quantum algorithms for GAD as quantum GAD and abbreviate it qGAD for convenience. In 2019, Liang et al.\cite{LSL19} proposed a qGAD algorithm, wherein the potential of exponential acceleration over the classical algorithm is highlighted, while its time complexity is not explicitly showcased. Subsequently, Guo et al.\cite{GLL22} pointed out the incorrect use of controlled rotation as a key subroutine in Liang's algorithm that would make it unworkable, and proposed a new qGAD algorithm based on amplitude estimation with exponential speedup in data size $M$ over the classical GAD. However, both existing quantum algorithms take phase estimation as key subroutines, the quantum circuits of which could have an exponentially long depth, making them resource-intensive and prone to decoherence and noise. This is unfavorable in the noisy intermediate-scale quantum (NISQ)\cite{NISQ18} and early fault-tolerant quantum eras. Moreover, both existing algorithms assume that the data provided is mean centered \cite{LSL19} or feature uncorrelated\cite{GLL22}, which may not hold in practice and thus limits their applicability.

To address these challenges, we in this paper further investigate qGAD, and propose a new qGAD algorithm without phase estimation. Specifically, our algorithm takes advantages of the technique of arithmetic-free quantum state preparation (AFQSP)\cite{QST19}, which originally aims to prepare a quantum state whose amplitudes is proportional to a real vector based on a black-box that returns the elements of the vector, to accomplish two new tasks: estimating the magnitude of the sum of a real vector and estimating the magnitude of the inner product between two real vectors. Armed with the capability of implementing such two tasks and given the quantum oracles having access to the elements of a training dataset, our algorithm estimates the magnitude of each element of the mean vector $\bm{\mu}$ and the covariance matrix $C$. To further reveal their signs, we develop a technique referred to as \textit{Hadamard sign test} that is to determine the sign of the product of two amplitudes in a qubit. This fulfills the core task of GAD as mentioned above, namely retrieving $\bm{\mu}$ and $C$. The quantum circuit of our algorithm has a depth only linear with the qubit count, which significantly reduces the burden of quantum hardware as apposed to the existing qGAD algorithms \cite{LSL19,GLL22}. In addition, our algorithm works for more generic data, namely relaxing the requirements of quantum form of data, input data mean centering and feature correlation in prior qGAD algorithms. We also show that the proposed quantum algorithm is exponentially faster than the classical GAD with respective to the data size $M$, and is also time competitive compared with the existing qGAD algorithms.

This paper is organized as follows. In section \ref{sec:2}, we review the classical GAD and AFQSP as two preliminaries of the proposed qGAD algorithm. In section \ref{sec:3}, we detail our algorithm. Its computational complexity is analyzed in section \ref{sec:4}. Hardware efficiency, time efficiency and data practicality of the proposed qGAD algorithm are discussed in section \ref{sec:5}. Finally, we conclude this paper in section \ref{sec:6}.

\section{Preliminaries}\label{sec:2}

Before detailing our quantum algorithm for GAD, we present two preliminaries as key materials or subroutines for better understanding the quantum algorithm. First, we review the classical GAD. Second, we briefly review arithmetic-free black-box quantum state preparation \cite{QST19}, which adopts quantum comparator\cite{RC04} as a key subroutine.

\subsection{Classical GAD}
\label{sec:CGAD}
GAD is a popular model-based approach for anomaly detection \cite{DM16}. Given a dataset composed of $M$ data points, $\mathcal{D}=\{\bm{x}_0, \cdots, \bm{x}_{M-1}\}$, and each one $\bm{x}_i=(x_{i0},\cdots, x_{iD-1})^\top \in \mathbb{R}^D$ is a $D$-dimensional column vector, so the whole dataset can be described by a \textit{data matrix} $X=(x_{ij})_{M \times D}$. 

In GAD, every data point $\bm{x}_i$ is assumed to be in an unknown $D$-dimensional Gaussian distribution, 
\begin{align}
    p(\bm{x})=\frac{1}{(2\pi)^{\frac{D}{2}}\abs{C}^{\frac{1}{2}}} e^{-\frac{1}{2}(\bm{x}-\bm{\mu})^\top C^{-1}(\bm{x}-\bm{\mu})},
    \label{eq:TargetProb}
\end{align}
which is characterized by the mean vector $\bm{\mu}$ and the covariance matrix $C$. Here, $\abs{C}$ stands for the determinant of $C$, $\bm{\mu}=\frac{\sum_{i=0}^{M-1} \bm{x}_i}{M} \equiv (\mu_0,\cdots, \mu_{D-1})$ with
\begin{align}
\label{eq:means}
    \mu_j&=\frac{\sum_{i=0}^{M-1} x_{ij}}{M},
\end{align}
and $C=\frac{\sum_{i=0}^{M-1} (\bm{x}_i-\bm{\mu})(\bm{x}_i-\bm{\mu})^\top}{M-1} =\frac{X(\mathbb{I}-\bm{e}\bm{e}^\top/M)X^\top}{M-1} \equiv (C_{jk})_{D\times D}$ with elements
\begin{align}
  \label{eq:covs}
      C_{jk}&= \frac{\sum_{i=0}^{M-1} (x_{ij}-\mu_j)(x_{ik}-\mu_k)}{M-1}\nonumber\\
            &=\frac{\sum_{i=0}^{M-1} x_{ij}x_{ik}}{M-1}-\frac{M\mu_j\mu_k}{M-1},
  \end{align}
where $\mathbb{I}$ is the identity matrix and $\bm{e}=(1,1,\cdots,1)^\top$.
Given a data point $\bm{x}'$, it is deemed to be an anomaly if $p(\bm{x}')$ is less than a predetermined threshold $\sigma$. To this end, the mean values $\bm{\mu}$ and the covariance matrix $C$ have to be acquired beforehand. Therefore, the ultimate task of GAD is to estimate $\bm{\mu}$ and $C$.

\subsection{AFQSP}
\label{sec:AFQSP}

AFQSP \cite{QST19} is a technique for quantum state preparation where data is transmitted from basis to amplitudes. Instead of doing arithmetic as required in the well-known Grover's black-box quantum state preparation algorithm \cite{Grover00}, AFQSP takes advantages of quantum comparator, which takes takes much less elementary gates and thus is more friendly on NISQ devices. Given a $M$-dimensional real-number vector $\bm{\alpha}=(\alpha_0,\alpha_1,\cdots,\alpha_{M-1})$ with each $\alpha_i$ for $i=0,1,\cdots, M-1$ satisfying $0 \le \alpha_i < 1$ and being assumed to be precise up to $n$ bits, AFQSP aims to prepare the target state
\begin{align}
\label{eq:targetstate}
\frac{1}{\norm{\bm{\alpha}}_2}\sum_{i=0}^{M-1} \alpha_i\ket{i}.   
\end{align} 
The arithmetic-free black-box quantum state preparation \cite{QST19} assumes we are provided a black-box oracle $O_{AMP}$ to access the amplitudes $\alpha_i$ that acts as 
\begin{align}
O_{AMP}: \ket{i}\ket{t}\mapsto \ket{i}\ket{t\oplus \alpha_i^{(n)}},
\end{align} 
where $t$ is a $n$-bit integer, $\oplus$ here is a bit-wise XOR operation, and  $\alpha_i^{(n)}=\lceil 2^n\alpha_i \rceil$. The process of arithmetic-free black-box quantum state preparation can be summarized by the following five steps.

(1) Initialize four quantum registers with each qubit in each register being in the zero state $\ket{0}$, and prepare the first register in the uniform superposition state by peforming a unitary operation denoted by $U_M$. $U_M$ acts as 
\begin{align}
U_M: \ket{0}^{\otimes \lceil\log_2(M)\rceil} \mapsto \frac{1}{\sqrt{M}}\sum_{i=0}^{M-1}\ket{i},
\end{align}
and is easy to implement; for example, if $M$ is two to the power of an integer $m$, i.e., $M=2^m$, $U_M$ would be the tensor product of $m$ Hadamard operations, that is, $U_M=H^{\otimes m}$. In general, $U_M$ can be implemented by using $O(\log(M))$ elementary one- and two-qubit gates with circuit depth $O(\log(M))$ \cite{SV24}. Thus the state of the four registers starts with 
\begin{align}
\label{state:AFQST1}
\frac{1}{\sqrt{M}}\sum_{i=0}^{M-1}\ket{i} \ket{0}^{\otimes n} \ket{0}^{\otimes n} \ket{0}.
\end{align}
The states of the first two registers represent the index of elements of the data vector $\bm{\alpha}$, the elements of $\bm{\alpha}$, so we term them \textit{index register} and \textit{data register} respectively.  The third register, called \textit{reference} register, is used for comparison, and the last register is referred to as the \textit{flag} register, respectively.

(2) Apply the oracle $O_{AMP}$ to the first two registers, yielding the state
\begin{align}
\frac{1}{\sqrt{M}}\sum_{i=0}^{M-1}\ket{i} \ket{\alpha_i^{(n)}} \ket{0}^{\otimes n} \ket{0}.
\end{align}

\begin{figure}[htb]
\centering
\includegraphics[scale=0.66]{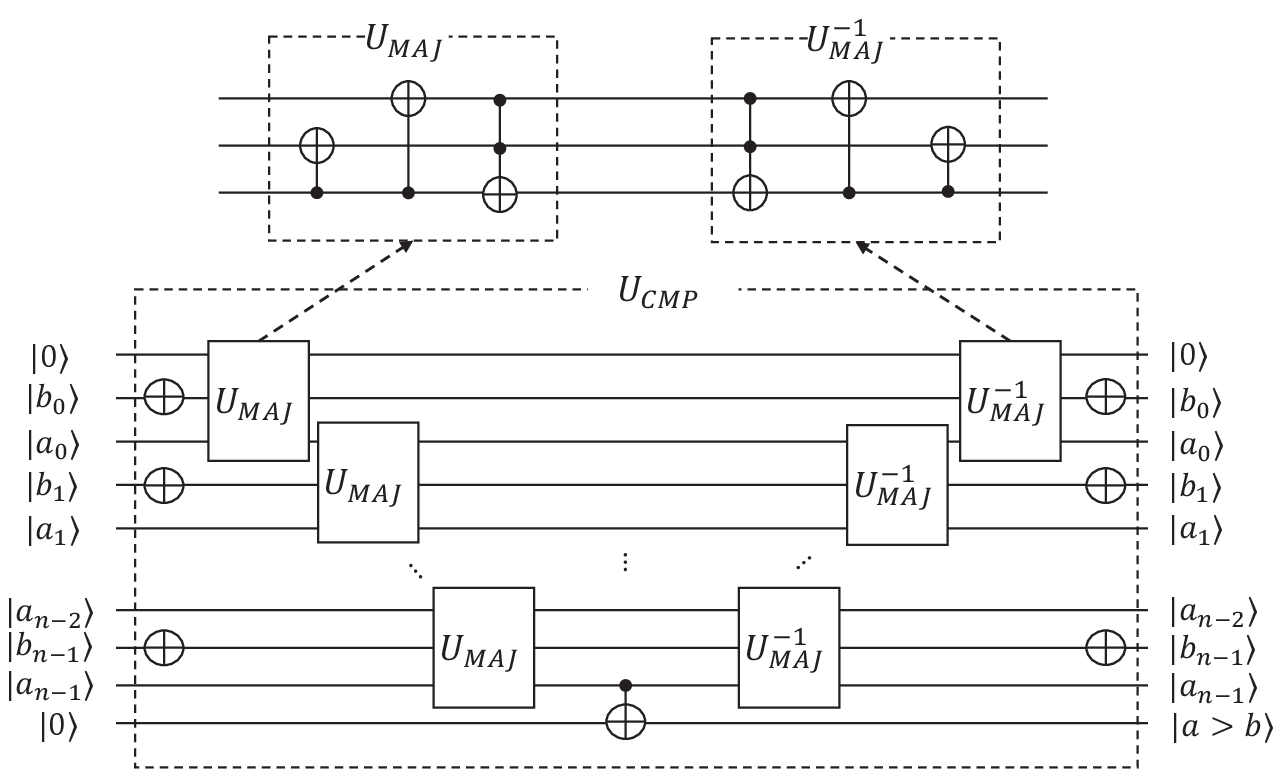}
\caption{Quantum circuit of $U_{CMP}$.} 
\label{fig:CMP}
\end{figure}

(3) Perform Hadamard gate on each qubit of the third register and then the quantum comparator \cite{RC04}, a unitary operation denoted by $U_{CMP}$, on the last three registers. $U_{CMP}$ acts as 
\begin{align}
\label{eq:CMP}
U_{CMP}: \ket{a}\ket{b}\ket{0} \mapsto \left\{ 
\begin{array}{rcl}
     \ket{a}\ket{b}\ket{1} &  &{a>b},  \\
     \ket{a}\ket{b}\ket{0} &  &{a\le b},
\end{array}
\right.   
\end{align} 
 and is mainly constructed by a sequence of  "MAJ" gates $U_{MAJ}$ and its inverse \cite{RC04}, whose quantum circuit can be shown in \ref{fig:CMP}. After this step, we now have the state
\begin{align}
\frac{1}{\sqrt{M2^n}}\sum_{i=0}^{M-1}\ket{i} \ket{\alpha_i^{(n)}}
\left(\sum_{x=0}^{\alpha_{i}^{(n)}-1}\ket{x} \ket{1} + \sum_{x=\alpha_{i}^{(n)}}^{2^n-1}\ket{x}\ket{0}\right).
\end{align}

(4) Again performing Hadamard gate on each qubit of the third register, we yield the state 
\begin{align}
\frac{1}{\sqrt{M}}\sum_{i=0}^{M-1} \ket{i} \ket{\alpha_i^{(n)}}\ket{0}^{\otimes n}\left(\frac{\alpha_i^{(n)}}{2^n} \ket{1} + \frac{2^n-\alpha_i^{(n)}}{2^n}\ket{0}\right)+  \ket{\omega},
\end{align}
where $\ket{\omega}$, and other states hereafter labeled by $\omega$ with or without subscripts, denote an unnormalized state containing the superposition of nonzero parts in the reference register, if not specified.

(5) Perform the inverse of $O_{AMP}$ to restore the state of the second register back to zero, and finally measure the last two registers in the state $\ket{0}^{\otimes n}\ket{1}$, then the target state \ref{eq:targetstate} is attained in the first register.

\begin{figure}[H]
\centering
\includegraphics[scale=0.6]{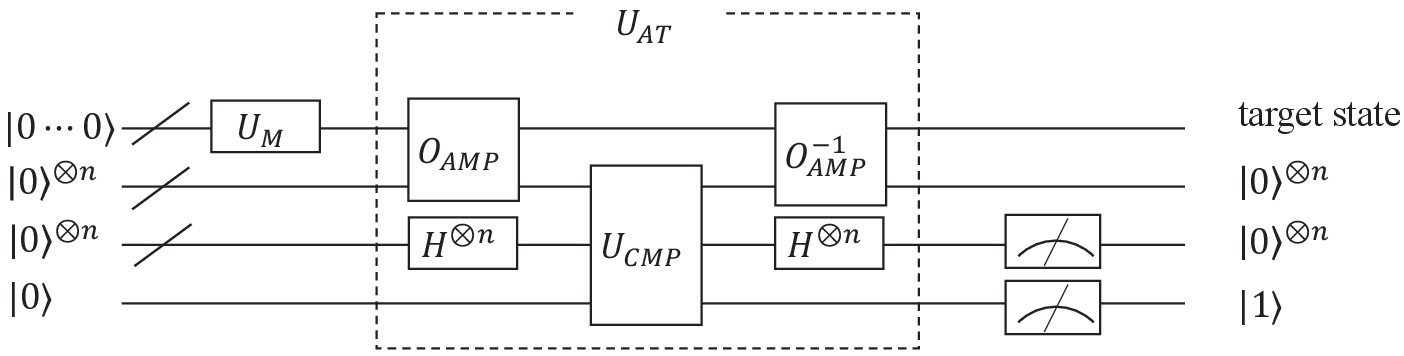}
\caption{Quantum circuit of AFQSP, where "/" in the wires indicates a bunch of qubits rather than only one qubit.} 
\label{fig:QST}
\end{figure}

The quantum circuit of the whole process is drawn in \ref{fig:QST}. As shown in \ref{fig:QST}, the central part of the circuit is constituted by the above steps (2)-(5) apart from the measurement, referred to as \textit{amplitude transduction} , which is a unitary operation denoted by $U_{AT}$ acting as
\begin{align}
\label{eq:UAT}
U_{AT}: \ket{i} \ket{0}^{\otimes n} \ket{0}^{\otimes n} \ket{0} \mapsto
\ket{i} \ket{0}^{\otimes n}\ket{0}^{\otimes n}\left(\frac{\alpha_i^{(n)}}{2^n} \ket{1} + \frac{2^n-\alpha_i^{(n)}}{2^n}\ket{0}\right)+  \ket{\omega_i}.
\end{align} 


\section{Quantum algorithm for GAD} \label{sec:3}
As mentioned in section \ref{sec:CGAD} and formulated in \eqref{eq:means} and \eqref{eq:covs}, the goal of GAD is to acquire the mean values  $(\mu_0,\cdots, \mu_{D-1})$ and the covariance matrix $C=(C_{jk})_{D\times D}$ of the multivariate Gaussian distribution used to model the given dataset $\mathcal{D}=\{\bm{x}_i=(x_{i0},\cdots,x_{iD-1})|i=0,\cdots,M-1\}$. So, our quantum algorithm consists of two parts: the first one aims to estimate the mean values $\mu_j$, while the second one works for estimating covariance matrix elements $C_{jk}$. As explained below, the two parts can be done in almost parallel, while estimating $C_{jk}$ requires the values of $\mu_j$ and $\mu_k$ as shown in \eqref{eq:covs}. 

Both parts of our quantum algorithm call for the access to the classical dataset $\mathcal{D}$ on a quantum computer. To accomplish it, we assume that quantum oracles that have access to $\mathcal{D}$ are provided. For convenience, each feature value $x_{ij}$ of each data point  
$\bm{x}_i$ is confined to $x_{ij}\in (-1,1)$, and is stored with precision up to $n$ bits, that is, $2^nx_{ij}$ has sign-and-magnitude binary form  $x_{ij}:=x_{ij}^{sgn}x_{ij}^{n-1}\cdots x_{ij}^{0}$ with $n+1$ bits, where the first bit $x_{ij}^{sgn}$ is a sign bit with $x_{ij}^{sgn}=0 (1)$ indicating non-negative (negative) $x_{ij}$ and the rest bits forms the absolute value of $x_{ij}$.  Our algorithms takes two kinds of oracles as follows,
\begin{align}
    \label{eq:qOracleSgn}
    O_j^{sgn}&: \ket{i}\ket{t}\mapsto \ket{i}\ket{t\oplus x_{ij}^{sgn}}, \\
    O_j^{abs}&: \ket{i}\ket{t}\mapsto \ket{i}\ket{t\oplus x_{ij}^{n-1}\cdots x_{ij}^{0}}, 
    \label{eq:qOracleAbs}
\end{align} 
for $j=1,2,\cdots,D$, which are to access the sign and the absolute value of $x_{ij}$ of all data points, respectively.
In our algorithm, $\ket{t}$ is set to be $\ket{0}$, so that $O_j^{sgn}$ and $O_j^{abs}$ can precisely load the feature values of all the data points in quantum parallel without being affected by the bit of $t$. One way of constructing such oracles is via quantum random access memory (QRAM) \cite{QRAM08,GLM08,HLGJ21}.  In this case, considering the size of $\mathcal{D}$, performing both $O_j^{sgn}$ and $O_j^{abs}$ would take time $O(\log(M))$\cite{QRAM08,GLM08,HLGJ21}; here we set $n=O(1)$ as its scaling does not depend on the size of $\mathcal{D}$. Alternatively, if every $x_{ij}$ is efficiently computable, the oracles can also be efficiently constructed.


\subsection{Part 1: Estimating the mean values}
\label{sec:EstMeans}

The first part of our quantum algorithm is to estimate all the mean values $\mu_1,\mu_2,\cdots,\mu_{D-1}$. For each $\mu_j$, our quantum algorithm runs the following steps to estimate it.

(1.1) Prepare five quantum registers with qubits being in the zero states, and then perform the unitary operation  $U_M$ on the first register to generate the uniform superposition state. This gives the state of all the registers 
\begin{align}
\label{state:EM1}
\frac{1}{\sqrt{M}}\sum_{i=0}^{M-1}\ket{i} \ket{0}\ket{0}^{\otimes n} \ket{0}^{\otimes n} \ket{0}.
\end{align} 
The state corresponds to the initial state \eqref{state:AFQST1} of AFQSP as described in subsection \ref{sec:AFQSP}, and the first, third, fourth and fifth register are also termed index register, data register, reference register and flag register for the same reason as stated in subsection \ref{sec:AFQSP}. The differences of these two states lie in that the state \eqref{state:EM1} introduces the second register with one qubit to store the sign of each feature value, called \textit{sign} register, due to signed feature values of our dataset.

(1.2) Apply the amplitude transduction $U_{AT}$ as shown in \eqref{eq:UAT} with $O_j^{abs}$ as the oracle to all the five registers except for the second, so we yield the state
\begin{align}
\label{state:xj}
\frac{1}{\sqrt{M}}\sum_{i=0}^{M-1} \ket{i} \ket{0}\ket{0}^{\otimes n}\ket{0}^{\otimes n}\left(\abs{x_{ij}} \ket{1} + (1-\abs{x_{ij}})\ket{0}\right)+  \ket{\omega_{11}}.
\end{align} 

(1.3) Perform the oracle $O_j^{sgn}$ on the first two registers to load the signs of $x_{ij}$ in the second register, the controlled-Z operation on the second register as the controller and the last register as the target to implement $\abs{x_{ij}}\mapsto x_{ij}$, and the inverse of oracle $O_j^{sgn}$ ($O_j^{sgn}$ itself) on the first two registers again to restore the state of the second register to zero. Now we have the state
\begin{align}
\label{state:AT}
\frac{1}{\sqrt{M}}\sum_{i=0}^{M-1} \ket{i} \ket{0}\ket{0}^{\otimes n}\ket{0}^{\otimes n}\left(x_{ij}\ket{1} + \left(1-\abs{x_{ij}}\right)\ket{0}\right)+  \ket{\omega_{12}}.
\end{align} 

The following two steps (1.4) and (1.4$^\prime$) are optional for estimating $\abs{\mu_j}$ and the sign of $\mu_j$, respectively.

(1.4) Perform the inverse of $U_M$ on the first register, and then measure the state of all the five registers to see the outcome $\ket{0}^{\otimes m}\ket{0}\ket{0}^{\otimes n}\ket{0}^{\otimes n} \ket{1}$. The success probability of measurement is 
\begin{align}
\label{eq:Pmu}
    P_{\mu}=\left(\frac{\sum_{i=0}^{M-1}x_{ij}}{M}\right)^2=\mu_j^2,
\end{align} 
which gives the estimate of $\abs{\mu_j}=\sqrt{P_{\mu}}$.

(1.4$^\prime$) Perform the inverse of $U_M$ on the first register, and then measure the state of the first four registers to see the outcome $\ket{0}^{\otimes m}\ket{0}\ket{0}^{\otimes n}\ket{0}^{\otimes n}$ with success probability 
\begin{align}
    P_{11}=\left(\frac{\sum_{i=0}^{M-1}x_{ij}}{M}\right)^2+\left(\frac{M-\sum_{i=0}^{M-1}\abs{x_{ij}}}{M}\right)^2.
\end{align}
The final state of the last flag qubit would be 
\begin{align}
\label{state:FlagFinal1}
\frac{1}{\sqrt{P_{11}}}\left(\frac{\sum_{i=0}^{M-1}x_{ij}}{M}\ket{1}+\frac{M-\sum_{i=0}^{M-1}\abs{x_{ij}}}{M}\ket{0}\right).
\end{align} 
Then we perform a Hadamard gate on this state, measure it to see the outcome $\ket{1}$, and the sign of $\mu_j$ is derived with a high probability as explained below.

Thanks to the use of AFQSP, the step (1.4$^\prime$) can acquire the sign of $\mu_j$ with a high probability. To see this, we present the technique we call \textit{Hadamard sign test} (\textbf{Theorem}~\ref{theorem:1}) to determine the sign of the product of two real amplitudes in a qubit state $\alpha\ket{1}+\beta\ket{0}$ with $\alpha\beta\neq 0$, i.e., the sign of $\alpha\beta$; the proof is provided in \ref{App:1}, and the procedure is simple: we just perform a Hadamard operation on this qubit state and measure it to see the outcome $\ket{1}$. Confining to the state \eqref{state:FlagFinal1} of the flag qubit in step (1.4$^\prime$), $\alpha=\frac{\sum_{i=0}^{M-1}x_{ij}}{\sqrt{P_{11}}M}=\frac{\mu_j}{\sqrt{P_{11}}}$ and $\beta=\frac{M-\sum_{i=0}^{M-1}\abs{x_{ij}}}{\sqrt{P_{11}}M}$. Since the size of each $x_{ij}$ and each $\abs{x_{ij}}$ generally does not depend on the number of data points $M$ and the number of features $D$, it is natural to assume each $x_{ij}, \abs{x_{ij}}\in \mathcal{O}(1)$, then $P_{11}=\mathcal{O}(1)$ and $\alpha,\beta=\mathcal{O}(1)$. Here, we assume the dataset $\mathcal{D}$ is not centered and the size of each mean values relies neither on $M$ nor $D$, that is, each $\mu_j\neq 0$ and $\mu_j \in \mathcal{O}(1)$. This is reasonable as if each $\mu_j=0$ or exponentially approaches to zero, there is no need to estimate the mean values and we can just set $\mu_j=0$ and then go straight to estimate the covariance matrix.   Therefore, according to \textbf{Theorem}~\ref{theorem:1}, the sign of $\alpha\beta$ can be correctly determined with a high probability taking only $\mathcal{O}(1)$ copies of the state \eqref{state:FlagFinal1}. Moreover, since $\abs{x_{ij}}<1$, $\beta>0$ always holds, thus the sign of $\alpha$ ($\mu_j$) is equal to that of $\alpha\beta$. Therefore, the sign of $\mu_j$ can be efficiently determined by taking $O(1)$ copies of the state \eqref{state:FlagFinal1}.

\theorem[Hadamard sign test]{  \textit{Given a normalized single-qubit state $\ket{\varphi}=\alpha\ket{1}+\beta \ket{0}$ with real amplitudes $\alpha$ and $\beta$ satisfying $\alpha^2+\beta^2=1$ and $\alpha\beta\neq 0$, there exists a procedure to determine the sign of $\alpha\beta$ with probability greater than or equal to $1-\delta$ using $\mathcal{O}\left(\frac{1}{\delta}\left(\frac{1}{\alpha^2}+\frac{1}{\beta^2}\right)\right)$ copies of $\ket{\varphi}$.}
\label{theorem:1}}

\begin{figure}[H]
\centering
\includegraphics[scale=0.66]{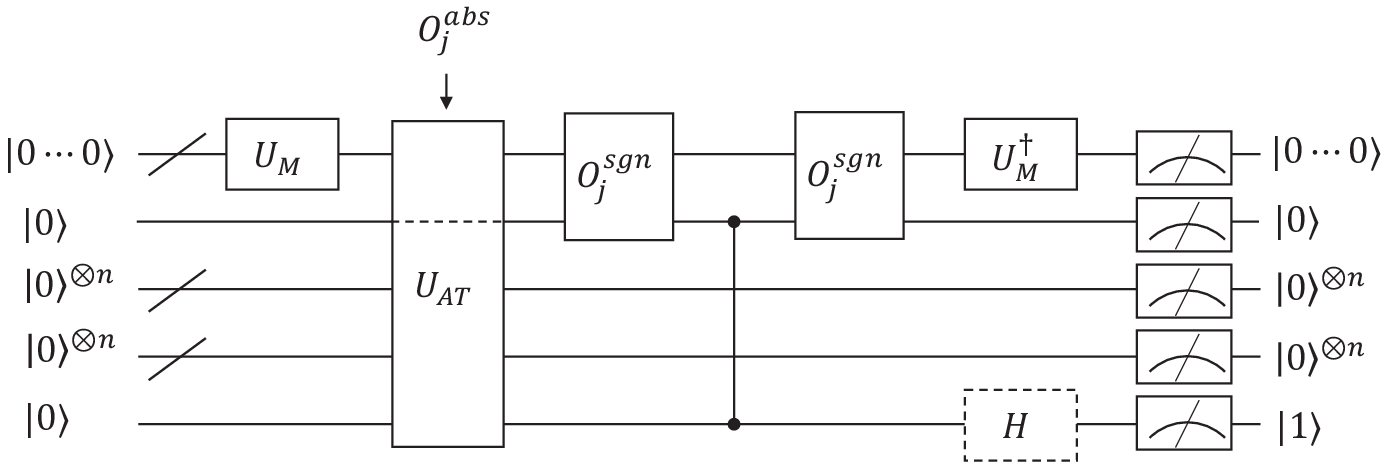}
\caption{Quantum circuit for estimating each mean value $\mu_j$, where $U_{AT}$ is the unitary operation of amplitude transduction defined in \eqref{eq:UAT} that acts on all the registers except for the second one; the dashed line in the $U_{AT}$ box means $U_{AT}$ does not act on the wire of the qubit in the second register. The dashed border line of the last Hadamard operation acting on the last qubit means that the Hadamard operation is operational; it is performed if the quantum circuit is to determine the sign of $\mu_j$, while it is not applied if the quantum circuit just runs for estimating the absolute value $\abs{\mu_j}$.} 
\label{fig:QEM}
\end{figure}

The quantum circuit of the whole procedure for estimating $\mu_j$, including its absolute value and its sign, is drawn in Figure \ref{fig:QEM}. 

It is worthy to note that we can alternatively directly estimate each $\mu_j$ instead of estimating $\abs{\mu_j}$ and the sign of $\mu_j$ separately, by implementing $x_{ij} \leftarrow x_{ij}^{\prime}:=(x_{ij}+1)/2$ which is always greater than zero as $x_{ij}\in (-1,1)$. By doing so, the mean value of  $x_{ij}^{\prime}$, denoted by  $\mu_j^{\prime}=\sum_{i=0}^{M-1}x_{ij}^{\prime}/M=(\mu_j+1)/2$, which is also greater than zero and can be directly estimated by the above algorithm in the step (1.4) without determining the sign by the step (1.4$^\prime$). This way, $\mu_j$ can be obtained as $\mu_j=2\mu_j^{\prime}-1$. However, implementing $x_{ij} \leftarrow x_{ij}^{\prime}:= (x_{ij}+1)/2$ entails adding a constant on a quantum computer, which would induces additional $O(n)$ qubits and $O(n)$ one- and two-qubit gates when resorting to ripple-carry quantum addition \cite{RC04}, or alternatively uses the quantum Fourier transformation based addition \cite{Draper00,RG17,Sahin20,Ferraz22} that would introduce additional $O(1)$ qubits and $O(n^2)$ one- and two-qubit gates. This is unfavorable on NISQ devices as much more quantum resources are required. Therefore, we prefer to estimate $\abs{\mu_j}$ and the sign of $\mu_j$ in separate.

\subsection{Part 2: Estimating the covariance matrix}
The second part of our quantum algorithm is to acquire the covariance matrix $C$ of the Gaussian distribution by estimating its each element $C_{jk}$.  Since the mean values $\mu_j$ are estimated by the first part of our algorithm, the second part only need focus on estimating each 
\begin{align}
    C_{jk}^\prime:=\frac{\sum_{i=0}^{M-1} x_{ij}x_{ik}}{M-1},
    \label{eq:Cjkprime}
\end{align} 
so that $C_{jk}$ can be estimated as $C_{jk}=C_{jk}^\prime-\frac{M\mu_j\mu_k}{M-1}$ according to the equation \eqref{eq:covs}. The following steps are run to estimate each $C_{jk}^\prime$.

(2.1) Prepare five quantum registers and run steps (1.1) and (1.2) to create the state \eqref{state:xj} .

(2.2) Measure the reference and flag registers to see the outcome $\ket{0}^{\otimes n}\ket{1}$, and then the state of the first register would be 
\begin{align}
\label{state:initialxj}
    \frac{1}{\sqrt{MP_{21}}}\sum_{i=0}^{M-1}\abs{x_{ij}}\ket{i},
\end{align}
where $P_{21}=\sum_{i=0}^{M-1}x_{ij}^2/M$ is the success probability of measurement.

(2.3) Perform Pauli X gate on the old flag qubit to recover its state from $\ket{1}$ to $\ket{0}$. Alternatively, we can discard the old reference and flag registers and prepare two new ones in the state $\ket{0}^{\otimes n}\ket{0}$. Then again apply $U_{AT}$ with oracle $O_k^{abs}$ to all the five registers except for the second one (i.e., the sign qubit). Now we have the state of the whole five registers
\begin{align}
\frac{1}{\sqrt{MP_{21}}}\sum_{i=0}^{M-1}\abs{x_{ij}} \ket{i} \ket{0}\ket{0}^{\otimes n}\ket{0}^{\otimes n}\left(\abs{x_{ik}} \ket{1} + \left(1-\abs{x_{ik}}\right)\ket{0}\right) +  \ket{\omega_{21}}.
\end{align} 

(2.4) Perform oracles $O_j^{sgn}$ and $O_k^{sgn}$ on the first two registers, controlled-Z operation on the second register, and oracles $O_k^{sgn}$ and $O_j^{sgn}$ again to load the signs of $x_{ij}$ and $x_{ik}$.  This gives the state

\begin{align}
\frac{1}{\sqrt{MP_{21}}}\sum_{i=0}^{M-1}\ket{i} \ket{0}\ket{0}^{\otimes n}\ket{0}^{\otimes n}\left(x_{ij}x_{ik} \ket{1} + \abs{x_{ij}}\left(1-\abs{x_{ik}}\right)\ket{0}\right) +  \ket{\omega_{22}}.
\end{align} 

Similar to the steps (1.4) and (1.4$^\prime$), the following steps (2.5) and (2.5$^\prime$) run in parallel to estimate $\abs{C_{jk}^\prime}$ and the sign of $C_{jk}^\prime$, respectively.

(2.5) Same as step (1.4), that is, performing $U_M^\dagger$ on the first register, and measuring the state of all the five registers to see $\ket{0}^{\otimes m}\ket{0}\ket{0}^{\otimes n}\ket{0}^{\otimes n} \ket{1}$. The success probability of measurement is 
\begin{align}
    P_{22}=\left(\frac{\sum_{i=0}^{M-1}x_{ij}x_{ik}}{M\sqrt{P_{21}}}\right)^2,
\end{align} 
which gives the estimate of $\abs{C^\prime_{jk}}$ as 
\begin{align}
    \abs{C^\prime_{jk}}=\frac{M\sqrt{P_{21}P_{22}}}{M-1}.
\end{align} 

(2.5$^\prime$) Same as step (1.5), namely perform $U_M^\dagger$ on the first register, and measure the state of the first four registers to see the state of zeros. The state of the last flag qubit would be  
\begin{align}
    \label{state:FlagFinanl2}
    \frac{1}{\sqrt{P_{23}}}\left(\frac{\sum_{i=0}^{M-1}x_{ij}x_{ik}}{M\sqrt{P_{21}}}\ket{1}+\frac{\sum_{i=0}^{M-1}\abs{x_{ij}}\left(1-\abs{x_{ik}}\right)}{M\sqrt{P_{21}}}\ket{0}\right),
\end{align} 
where 
\begin{align}
P_{23}=\left(\frac{\sum_{i=0}^{M-1}x_{ij}x_{ik}}{M\sqrt{P_{21}}}\right)^2+\left(\frac{\sum_{i=0}^{M-1}\abs{x_{ij}}(1-\abs{x_{ik}})}{M\sqrt{P_{21}}}\right)^2
\end{align} 
represents the success probability of the measurement. Then, as used in the step (1.4$^\prime$), the Hadmard sign test is applied to retrieve the sign of  $C^\prime_{jk}$. Note that in the state \eqref{state:FlagFinanl2}, the sign of the amplitude of $\ket{0}$ is positive, and the sign of the amplitude of $\ket{1}$ coincides with that of $C^\prime_{jk}$. Thus, according to \textbf{Theorem}~\ref{theorem:1}, the sign of $C^\prime_{jk}$ can be determined with a high probability.
It is worthy to note that, both the matrix $C'=(C_{jk}^\prime)$ and the covariance matrix $C$ are symmetric, and thus only $\frac{D(D+1)}{2}$ rather than $O(D^2)$ elements of $C'$ and $C$ are needed to estimate.

\begin{figure*}[htb]
\centering
\includegraphics[scale=0.52]{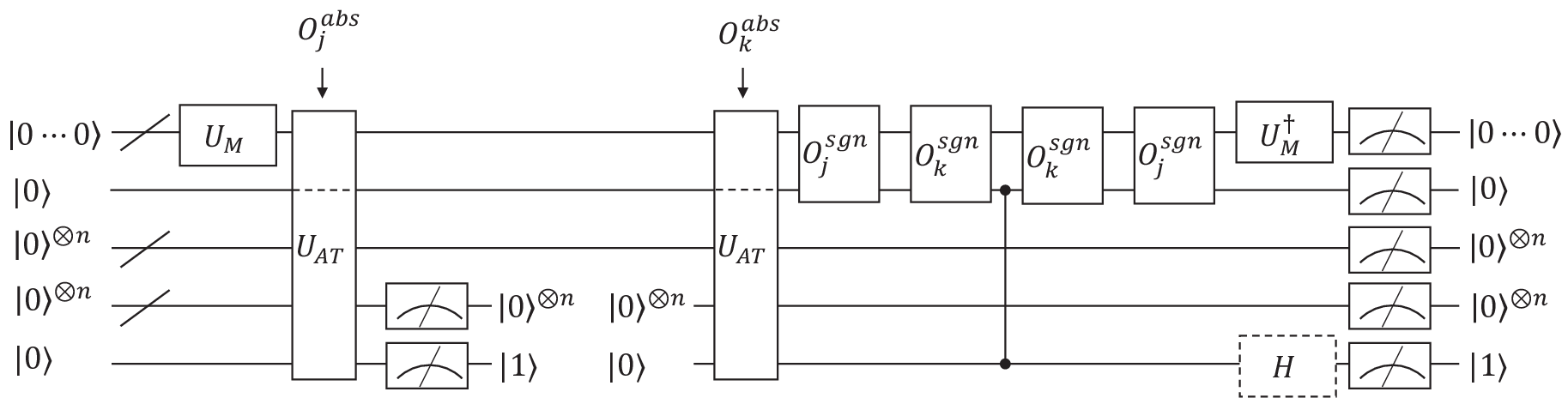}
\caption{Quantum circuit for estimating each element $C^\prime_{jk}$, where the first $U_{AT}$ takes $O_j^{abs}$ as the oracle, while the second one takes oracle $O_k^{abs}$. Just as shown in Figure \ref{fig:QEM}, the Hadamard operation in the dashed box is applied if this circuit is utilized for retrieving the sign of $C^\prime_{jk}$, while it is not applied if the circuit works for estimating $\abs{C^\prime_{jk}}$.} 
\label{fig:QECOV}
\end{figure*}

The whole quantum circuit for estimating each $C^\prime_{jk}$ is drawn in Figure \ref{fig:QECOV}.

\section{Complexity analysis}\label{sec:4}
In this section, we individually analyze the time complexity of two parts of our quantum algorithm, and conclude the total time complexity of the whole algorithm. 

\subsection{Time complexity of estimating mean values}
\label{sec:TimeEstMean}
We first analyze the time complexity of estimating $\mu_j(j=1,2,\cdots D)$ following the steps (1.1)-(1.4$^\prime$) described in section~\ref{sec:EstMeans}. Step (1.1) involves performing $U_M$ which takes time $O[\log(M)]$ as shown in subsection~\ref{sec:AFQSP}. Step (1.2) performs amplitude transduction unitary operation $U_{AT}$ whose quantum circuit is drawn in Figure~\ref{fig:QST}, whereas the oracle $O_{AMP}$ in the circuit should be replaced by $O_j^{abs}$. Considering $O_j^{abs}$ takes time $O[\log(M)]$ and $n=O(1)$ as analyzed above, $U_{AT}$ takes time $O[\log(M)]$. Putting these three together, steps (1.1)-(1.3) takes time $O[\log(M)]$  to generate one copy of the state \eqref{state:AT}.

As for the step (1.4) for estimating $\abs{\mu_j}$, to ensure $\abs{\mu_j}$ can be estimated within error $\epsilon_\mu$, the error of estimating $P_\mu$ has to be within $O\left(\sqrt{P_\mu}\epsilon_\mu\right)$ according to the equation \eqref{eq:Pmu}, which results in $O\left[\frac{P_\mu(1-P_\mu)}{P_\mu\epsilon_\mu^2}\right]=O\left(\frac{1}{\epsilon_\mu^2}\right)$ copies of state \eqref{state:AT} are required for measurement in the step (1.4). Thus, $\abs{\mu_j}$ can be estimated within error $\epsilon_\mu$ in time $O\left[\frac{\log(M)}{\epsilon_\mu^2}\right]$. It is worthy to note that, the value of $\abs{\mu_j}$ significantly less than $\epsilon_\mu$ would be estimated to be zero and is unnecessary to proceed to estimate its sign, so only $\abs{\mu_j}=\Omega(\epsilon_\mu)$ matters in the step (1.4$^\prime$).

As for the steps (1.1)-(1.3) and (1.4$^\prime$) to estimate the sign of $\mu_j$, the sign can be correctly determined with probability $\ge 1-\delta$ using $O\left(\frac{1}{\alpha^2\beta^2\delta}\right)$ copies of the state \eqref{state:FlagFinal1} according to \textbf{Theorem}~\ref{theorem:1}. Here,
\begin{align}
\alpha^2=\frac{1}{P_{11}}\left(\frac{\sum_{i=0}^{M-1}x_{ij}}{M}\right)^2 
= \frac{\mu_j^2}{P_{11}} 
=\Omega\left(\frac{\epsilon_\mu^2}{P_{11}}\right)
\end{align} 
and 
\begin{align}
\beta^2=\frac{1}{P_{11}}\left(\frac{M-\sum_{i=0}^{M-1}\abs{x_{ij}}}{M}\right)^2\ge \frac{1}{4^nP_{11}},
\end{align} 
due to the fact that every $x_{ij}$ lies in (-1,1) with precision up to $n$ bits and $\abs{x_{ij}}\le 1-\frac{1}{2^n}$. This means that $O\left[\frac{P_{11}}{\delta}\left(\frac{1}{\epsilon_\mu^2}+4^n\right)\right]$ copies of the state \eqref{state:FlagFinal1}, and hence $O\left[\frac{1}{P_{11}}\frac{P_{11}}{\delta}\left(\frac{1}{\epsilon_\mu^2}+4^n\right)\right] =O\left[\frac{1}{\delta}\left(\frac{1}{\epsilon_\mu^2}+4^n\right)\right]$  copies of the state \eqref{state:AT}, are required. Therefore, the sign of $\mu_j$ can be determined with probability $\ge 1-\delta$ in time $O\left[\frac{\log(M)}{\delta}\left(\frac{1}{\epsilon_\mu^2}+4^n\right)\right] =O\left[\frac{\log(M)}{\epsilon_\mu^2}\right]$ due to $n,\delta=O(1)$.

Putting above results together, we can conclude that estimating all $D$ mean values $\mu_1,\mu_2,\cdots,\mu_D$ has time complexity $O\left[\frac{D\log(M)}{\epsilon_\mu^2}\right]$.

\subsection{Time complexity of estimating covariance matrix}

The steps (2.1)-(2.5$^\prime$) , whose quantum circuit is drawn in Figure \ref{fig:QECOV}, are executed to estimate all values $C_{jk}^{\prime}(j,k=1,2,\cdots,D)$, and further estimate all the elements $C_{jk}(j,k=1,2,\cdots,D)$ of the covariance matrix $C$ according to the equation \eqref{eq:Cjkprime}. Step (2.1) implements $U_M$ and $U_{AT}$, both of which evidently take time $O[\log(M)]$ as analyzed in subsection~\ref{sec:TimeEstMean}. In the step (2.2),suppose $\mathcal{N}$ times of measurement are performed, and consequently $P_{21}$ can be  estimated within error 
$$O\left[\frac{\sqrt{P_{21}(1-P_{21})}}{\sqrt{\mathcal{N}}}\right],$$ and roughly $\mathcal{N}P_{21}$ copies of the state \eqref{state:initialxj} would be output. Considering $U_{AT}$,$U_{M}^\ddag$, and oracles involved all take time $O[\log(M)]$, the steps (2.3) and (2.4) take time $O[\log(M)]$. 
In the step (2.5), $P_{22}$ would be estimated with error 
$$O\left[\frac{\sqrt{P_{22}(1-P_{22})}}{\sqrt{\mathcal{N}}P_{21}}\right].$$
This results in the error of estimating $\abs{C_{jk}^{\prime}}$ scaling as 
\begin{align}
O\left[\frac{\sqrt{P_{22}}}{\sqrt{P_{21}}}\frac{\sqrt{P_{21}(1-P_{21})}}{\sqrt{\mathcal{N}}}+\frac{\sqrt{P_{21}}}{\sqrt{P_{22}}}\frac{\sqrt{P_{22}(1-P_{22})}}{\sqrt{\mathcal{N}P_{21}}}\right]
=&O\left[\frac{\sqrt{P_{22}(1-P_{21})}+\sqrt{1-P_{22}}}{\sqrt{\mathcal{N}}}\right]\\
=&O\left(\frac{1}{\sqrt{\mathcal{N}}}\right).
\end{align}
Thus, to make the error of estimating $\abs{C_{jk}^\prime}$ also be within $O\epsilon_\mu$, $\mathcal{N}=O\left(\frac{1}{\epsilon_\mu^2}\right)$ copies of state~\eqref{state:xj} are required.  This would makes the error of estimating $C_{jk}$ that we denote by $\epsilon_C$ less than $3\epsilon$ according to the equation~\eqref{eq:Cjkprime}. So the total time for estimating the absolute values of all the elements (i.e., $\abs{C_{jk}}$) of the covariance matrix $C$ within error $\mathcal{O}(\epsilon_\mu)$ scales as $\mathcal{O}\left[\frac{D^2\log(M)}{\epsilon_\mu^2}\right]$, provided that we have already acquired the the estimates of all $\mu_j$ and that the time for preparing the state \eqref{state:xj} scales as $\mathcal{O}(\log(M))$ as shown in subsection~\ref{sec:TimeEstMean}.

Now we analyze the time complexity for estimating the sign of $C_{jk}^\prime$. Similar to estimating the sign of mean values $\mu_j$, only if the magnitude of $C_{jk}^\prime$ is significantly large, i.e., $\abs{C_{jk}^{\prime}}=\Omega(\epsilon_\mu)$, it is necessary to run the steps (2.1)-(2.4) and (2.5$^\prime$) to estimate the sign of $C_{jk}^{\prime}$. As shown in the step (2.5$^\prime$), according to \textbf{Theorem}~\ref{theorem:1}, $O\left[\frac{1}{\delta}(\frac{1}{\alpha^2}+\frac{1}{\beta^2})\right]$ copies of the state \eqref{state:FlagFinanl2} are taken to ensure the sign of $C_{jk}^{\prime}$ can be correctly determined with probability $\ge 1-\delta$. Here 
\begin{align}
    \alpha^2=\frac{1}{P_{21}P_{23}}\left(\frac{\sum_{i=0}^{M-1}x_{ij}x_{ik}}{M}\right)^2
    \approx \frac{C_{jk}^{\prime 2}}{P_{21}P_{23}}=\Omega\left(\frac{\epsilon_\mu^2}{P_{21}P_{23}}\right)
\end{align} 
and 
\begin{align}
\beta^2&=\frac{1}{P_{21}P_{23}}\left(\frac{\sum_{i=0}^{M-1}\abs{x_{ij}}(1-\abs{x_{ik}})}{M}\right)^2 \nonumber\\
&\ge \frac{1}{P_{21}P_{23}}\left(\frac{\sum_{i=0}^{M-1}\abs{x_{ij}}}{2^nM}\right)^2  \nonumber\\ 
&\ge \frac{\mu_j^2}{P_{21}P_{23}4^n}
=\Omega\left(\frac{\epsilon_\mu^2}{P_{21}P_{23}4^n}\right).
\end{align} 
As a result, $O\left[\frac{P_{21}P_{23}}{\delta}\left(\frac{1}{\epsilon_\mu^2}+\frac{4^n}{\epsilon_\mu^2}\right)\right]$ copies of the state \eqref{state:FlagFinanl2}, and thus also $O\left[\frac{1}{\delta}\left(\frac{1}{\epsilon_\mu^2}+\frac{4^n}{\epsilon_\mu^2}\right)\right]=O\left[\frac{1}{\epsilon_\mu^2}\right]$ copies of the state \eqref{state:xj} in the step (2.1), are sufficient. Therefore, the time for estimating the sign of each $C_{jk}^{\prime}$ scales as $O\left[\frac{\log(M)}{\epsilon_\mu^2}\right]$, which is same as that of  estimating all $\abs{C_{jk}^{\prime}}$, and gives the overall time complexity for estimating all the elements of the covariance matrix $C$ being $O\left[\frac{D^2\log(M)}{\epsilon_\mu^2}\right]$. It is not hard to see that the time for estimating the covariance matrix $C$ is much larger than that for estimating all $D$ mean values $\bm{\mu}$, and thus the whole time complexity of the proposed quantum algorithm for GAD is 
\begin{align}
    \mathcal{O}\left(\frac{D^2\log(M)}{\epsilon_\mu^2}\right).
    \label{eq:TimeWholeMu}
\end{align}

\subsection{Time complexity of the whole algorithm}

Keeping in mind that the ultimate goal of our quantum algorithm is to estimate the probability $p(\bm{x})$ (equation~\eqref{eq:TargetProb}), the overall time complexity of our quantum algorithm depends on the error of estimating $p(\bm{x})$.  Hence, it is necessary to see how this error is affected by the errors of estimating  $\mu_j$ and $C_{jk}$, i.e., $\epsilon_\mu$ and $\epsilon_C$. For convenience, we use  $\Delta Q :=\hat{Q}-Q$ to denote the difference between the estimate of a quantity $Q$ denoted by $\hat{Q}$; the error of estimating $p(\bm{x})$ would be $\abs{\Delta p(\bm{x})}$ . Following this definition, it is easy to see $\Delta\bm{\mu}=\hat{\bm{\mu}}-\bm{\mu}$ would be a column vector with elements being $\hat{\mu}_j-\mu_j\le \epsilon_\mu$ and  $\Delta C= \hat{C}-C$  would be a matrix with elements being $\hat{C}_{jk}-C_{jk}\le \epsilon_C$. According to the equation~\eqref{eq:TargetProb},  $\abs{\Delta p(\bm{x})}$ is upper bounded by 
 \begin{align}
 \abs{\Delta p(\bm{x})} &=\abs{\Delta(e^{\ln p(\bm{x})})} \\
 &\approx p(\bm{x})\abs{\Delta(\ln p(\bm{x}))} \\
 &\le \abs{\Delta(\ln p(\bm{x}))}\\
 &=\frac{1}{2}\abs{\Delta(\ln \abs{C})+\Delta\left((\bm{x}-\bm{\mu})^\top C^{-1} (\bm{x}-\bm{\mu})\right)}\\
 &\le \frac{1}{2}\abs{\Delta(\ln \abs{C})}+\frac{1}{2}\abs{\Delta\left((\bm{x}-\bm{\mu})^\top C^{-1} (\bm{x}-\bm{\mu})\right)}.
 \label{ieq:Deltapx}
 \end{align}
Next, we individually analyze the upper bound of $\abs{\Delta(\ln \abs{C})}$ and that of $\abs{\Delta\left((\bm{x}-\bm{\mu})^\top C^{-1} (\bm{x}-\bm{\mu})\right)}$. According to the determinant relative error bound \cite{IR08}, the former one can be derived as 
\begin{align}
  \abs{\Delta(\ln \abs{C})} &\approx \abs{\frac{ \abs{\hat{C}}-\abs{C}}{\abs{C}}}\\
  &\le \frac{D\norm{C^{-1}} \norm{\Delta C}}{1-D\norm{C^{-1}} \norm{\Delta C}}\\
  &\le 2D\norm{C^{-1}}\norm{\Delta C},
  \label{ieq:DeltaLnDetC}
\end{align}
provided that $D\norm{C^{-1}}\norm{\Delta C}\le 1/2<1$. Here $\norm{\bullet}$ stands for the 2-norm of a vector or the $L_2$ norm of a matrix. Letting $\bm{y}:=\bm{x}-\bm{\mu}$ and $\hat{\bm{y}}:=\bm{x}-\hat{\bm{\mu}}$, we have
\begin{align}
    \abs{\Delta\left((\bm{x}-\bm{\mu})^\top C^{-1} (\bm{x}-\bm{\mu})\right)}
    &=\abs{(\bm{x}-\bm{\mu})^\top C^{-1} (\bm{x}-\bm{\mu})-(\bm{x}-\hat{\bm{\mu}})^\top \hat{C}^{-1} (\bm{x}-\hat{\bm{\mu}})}\\
    &=\abs{\bm{y}^\top C^{-1} \bm{y}-\hat{\bm{y}}^\top \hat{C}^{-1} \hat{\bm{y}}}\\
    &=\abs{\bm{y}^\top C^{-1} \bm{y}- \bm{y}^\top \hat{C}^{-1} \hat{\bm{y}}+\bm{y}^\top \hat{C}^{-1} \hat{\bm{y}}-\hat{\bm{y}}^\top \hat{C}^{-1} \hat{\bm{y}}}\\
    &\le \abs{\bm{y}^\top C^{-1} \bm{y}- \bm{y}^\top \hat{C}^{-1} \hat{\bm{y}}}+\abs{\bm{y}^\top \hat{C}^{-1} \hat{\bm{y}}-\hat{\bm{y}}^\top \hat{C}^{-1} \hat{\bm{y}}}\\
    &\le \norm{\bm{y}}\norm{C^{-1} \bm{y}-\hat{C}^{-1} \hat{\bm{y}}}+\norm{\bm{y}-\hat{\bm{y}}}\norm{\hat{C}^{-1}\hat{\bm{y}}}\\
    &\le \norm{\bm{y}}\norm{\hat{C}^{-1}}(\norm{\hat{\bm{y}}-\bm{y}}+\norm{\Delta C}\norm{ C^{-1}}\norm{\bm{y}})+\norm{\bm{y}-\hat{\bm{y}}}\norm{\hat{C}^{-1}}\norm{\hat{\bm{y}}}\\
    \label{ieq:InverseDeltax}
    &= \norm{\hat{C}^{-1}}\norm{\hat{\bm{y}}-\bm{y}}(\norm{\bm{y}}+\norm{\hat{\bm{y}}})+\norm{ \hat{C}^{-1}}\norm{ C^{-1}}\norm{\Delta C}\norm{\bm{y}}^2 \\
    &\le 2\norm{C^{-1}}\norm{\hat{\bm{y}}-\bm{y}}(\norm{\bm{y}}+\norm{\hat{\bm{y}}})+2\norm{C^{-1}}^2\norm{\Delta C}\norm{\bm{y}}^2,
    \label{ieq:InverseHatC}
\end{align}
where the above inequality~\eqref{ieq:InverseDeltax} holds due to the fact that $\hat{C}^{-1} \hat{\bm{y}}-C^{-1} \bm{y}=\hat{C}^{-1} (\hat{\bm{y}}-\bm{y}-\Delta C C^{-1} \bm{y})$ , and the inequality~\eqref{ieq:InverseHatC} holds as $\norm{\hat{C}^{-1}}\le \norm{C^{-1}}/(1-\norm{\Delta C C^{-1}})\le 2\norm{C^{-1}}$ for 
$\norm{\Delta C C^{-1}}\le \norm{C^{-1}}\norm{\Delta C} \le D\norm{C^{-1}}\norm{\Delta C} \le 1/2$; the results can be seen in \cite{Higham02}.

Putting the inequalities~\eqref{ieq:Deltapx},\eqref{ieq:DeltaLnDetC} and \eqref{ieq:InverseHatC} together, we have 
\begin{align}
    \abs{\Delta p(\bm{x})} &\le  D\norm{C^{-1}}\norm{\Delta C}+\norm{C^{-1}}\norm{\hat{\bm{y}}-\bm{y}}(\norm{\bm{y}}+\norm{\hat{\bm{y}}})+\norm{C^{-1}}^2\norm{\Delta C}\norm{\bm{y}}^2\\
    &\le D^2\norm{C^{-1}}\epsilon_C+4D\norm{C^{-1}}\epsilon_\mu+4D^2\norm{C^{-1}}^2\epsilon_C
    \label{ieq:ErrorDeltaCDeltay}\\
    &\le 3D^2\norm{C^{-1}}\epsilon_\mu+4D\norm{C^{-1}}\epsilon_\mu+12D^2\norm{C^{-1}}^2\epsilon_\mu
    \label{ieq:ErrorCtoErrormu}\\
    &\le 7D^2\norm{C^{-1}}\epsilon_\mu+12D^2\norm{C^{-1}}^2\epsilon_\mu.
    \label{ieq:UpperboundLower}
\end{align}
Here the inequality~\eqref{ieq:ErrorDeltaCDeltay} holds as the absolute value of each element of $\Delta C$ is within $\epsilon_C$ and $\norm{\Delta C}\le D\epsilon_C$, the absolute value of each element of $\hat{\bm{y}}-\bm{y}$ is within $\epsilon_\mu$ and $\norm{\hat{\bm{y}}-\bm{y}}\le \sqrt{D}\epsilon_\mu$, and each element of $\bm{y}$ and $\hat{\bm{y}}$ lies in $(-2,2)$ and thus $\norm{\hat{\bm{y}}},\norm{\bm{y}} \le 2\sqrt{D}$. The inequality~\eqref{ieq:ErrorCtoErrormu} holds due to $\epsilon_C\le 3\epsilon_\mu$ as analyzed above, and the inequality~\eqref{ieq:UpperboundLower} is derived for $4D\norm{C^{-1}}\epsilon_\mu\le 4D^2\norm{C^{-1}}\epsilon_\mu$. In order to make $p(\bm{x})$ be estimated within error $\epsilon$, i.e., $\abs{\Delta p(\bm{x})}\le \epsilon$, we set $\epsilon_\mu=\epsilon/(7D^2\norm{C^{-1}}+12D^2\norm{C^{-1}}^2)$ as implied by the inequality~\eqref{ieq:UpperboundLower}. Furthermore, since $\norm{C}$ corresponds to the largest eigenvalue for the positive definite matrix $C$ and satisfy $\norm{C}=\norm{\frac{X(\mathbb{I}-\ketbra{\bm{e}}{\bm{e}})X^\top}{M-1} \equiv (C_{jk})_{D\times D}} \le \frac{\norm{X}\norm{\mathbb{I}-\ketbra{\bm{e}}{\bm{e}}}\norm{X^\top}}{M-1} = \frac{\norm{X}\norm{X^\top}}{M-1} \le D$, we let the eigenvalues of $C$ lie in $[D/\kappa,D]$, where $\kappa$ is an effective condition number of $C$ that upper bounds its actual condition number. Consequently, $\norm{C^{-1}}\le \kappa/D$, 
\begin{align}
    \frac{1}{\epsilon_\mu}\le \frac{7D\kappa+12\kappa^2}{\epsilon},
\end{align}
and the overall time complexity of our quantum algorithm for estimating $p(\bm{x})$ within error $\epsilon$ according to the equation~\eqref{eq:TimeWholeMu} is 
\begin{align}
    \mathcal{O}\left(\frac{D^2\kappa^2(D+\kappa)^2\log(M)}{\epsilon^2}\right).
\end{align}

\section{Discussion}\label{sec:5}

In this section, we discuss the advantages of the proposed qGAD algorithm over the classical GAD algorithm and the prior qGAD algorithms. Comparison among the three kinds of GAD algorithms is listed in the Table~\ref{tab1}, from which we conclude advantages of the proposed qGAD algorithm from the aspects of hardware efficiency, time efficiency, and data practicality shown below. 

\subsection{Hardware efficiency}
Previous qGAD algorithms \cite{LSL19,GLL22} takes amplitude estimation as key subroutines so that their quantum circuit depth and size is exponential with the number of qubits up to which estimation has precision. As for the qGAD algorithm \cite{GLL22} which takes the essentially same oracles as ours in equations \eqref{eq:qOracleAbs} and \eqref{eq:qOracleSgn}, it has quantum circuit depth and size scaling as $O(2^nd)$ and $O(2^n s)$, where $n$ is the number of bits of estimation precision in amplitude estimation, $d$ and $s$ respectively represent the circuit depth and size of controlled oracles. By contrast, as shown in Figures \ref{fig:QEM} and \ref{fig:QECOV}, our algorithm significantly reduce the circuit depth and size to $O(d+n)$ and $O(s+n)$ respectively. Since shallower quantum circuit depth and smaller circuit size means less opportunity for errors to occur, the proposed algorithm is relatively more hardware efficient than the existing qGAD algorithms.

\subsection{Time efficiency}

From Table~\ref{tab1}, it is easy to see the proposed qGAD algorithm has time complexity $\mathcal{O}\left(\polylog{M}\right)$ when $D,\kappa,1/\epsilon=\mathcal{O}(\polylog{M})$, while the classical GAD algorithm takes time complexity $\mathcal{O}(M\polylog{M})$  under the same conditions. Hence, the proposed qGAD algorithm has potential to achieve exponential speedup over the classical GAD. However, when comparing the proposed qGAD algorithm with existing qGAD algorithms \cite{LSL19,GLL22} , it is unfair to do direct comparison because their current time complexity are derived under different assumptions on input data. Due to the incorrectness of the qGAD algorithm \cite{LSL19} as shown in \cite{GLL22}, we only consider the qGAD algorithm \cite{GLL22} which assume the data features are mutually uncorrelated and the covariance matrix is thus diagonal. Under the same assumption, the proposed algorithm would have time complexity  $\mathcal{O}\left(D\kappa^2(D+\kappa)^2\log(M)/\epsilon^2\right)$ with better dependence on $\epsilon$ but worse on $D$ and $\kappa$ when compared with the qGAD algorithm \cite{GLL22}. 

One may concern if the proposed qGAD algorithm can be dequantized by the technique of quantum-inspired classical sampling, which has made many QML or QML related algorithms lose exponential speedups over their classical counterparts, such as quantum algorithm for recommendation systems \cite{Tang19}, quantum principal component analysis \cite{Tang21}, quantum clustering \cite{Tang21}, and quantum linear regression \cite{GST22}.  An overview of quantum-inspired classical algorithms can be seen in \cite{Tang22,Tang23}. Given a dataset of data points (vectors) $\{\mathbf{x}_i|i=0,1,\cdots,M-1\}$, most existing QML algorithms assume there is a efficient procedure to create $\ket{\mathbf{x}_i}$ for every data vector $\mathbf{x}_i$ and the states of all the data points $\ket{\mathbf{x}_i}$ together with  the norms $\norm{\mathbf{x}_i}$ can be efficiently accessed in quantum parallel via quantum oracles. The quantum oracles can be implemented by QRAM provided a special data structure in which subsets of sums of squared data vector entries are stored in a binary tree \cite{Tang19,Tang23}. To make the classical counterparts more comparable to QML algorithms, quantum-inspired algorithms assume an analogous and reasonable input data model, i.e., sampling and query (SQ) access to a vector or a matrix. More specifically, $\mathrm{SQ}(\mathbf{v})$, sampling and query access to a column vector $\mathbf{v}=(v_0,v_1,\cdots,v_{M-1})^\top\in \mathbb{C}^M$, supports three efficient operations: (\romannumeral 1) querying for $v_i$ for all $i=0,1,\cdots,D-1$; (\romannumeral 2) sampling $v_i$ with probability $v_i^2/\norm{\mathbf{v}}^2$; (\romannumeral 3) querying for $\norm{\mathbf{v}}$. The SQ model allows efficient estimation of inner product, singular value transformation, matrix-vector multiplication and so on \cite{Tang23}. Let us take inner product estimation as an example.
Given $\mathrm{SQ}(\mathbf{u})$ and $\mathrm{SQ}(\mathbf{v})$ for two vectors $\mathbf{u},\mathbf{v}\in \mathbb{C}^M$, one can efficiently estimate the inner product $\mathbf{u}^\top\mathbf{v}$ within additive error $\norm{\mathbf{u}}\norm{\mathbf{v}}\epsilon$ with at least $1-\delta$ probability using only $\mathcal{O}\left(\frac{1}{\epsilon^2}\log\left(\frac{1}{\delta}\right)\right)$ queries and samplings, which is independent of the dimensionality $M$. 

As for our qGAD algorithm, efficient estimating inner product is crucial because each mean value $\mu_j=\mathbf{e}^\top\mathbf{x}_{\cdot j}/M$ and each $C^{\prime}_{jk}=\mathbf{x}_{\cdot j}^\top\mathbf{x}_{\cdot k}$ can be seen as the inner product of two vectors, where $\mathbf{e}$ is a $D$-dimensional vector with every element being one, and $\mathbf{x}_{\cdot j}=(x_{0j},x_{1j},\cdots,x_{M-1j})$ represents a column vector of $(j+1)$th feature values of all $M$ data points. If we have $\mathrm{SQ}(\mathbf{x}_{\cdot j})$ for all $j=0,1,\cdots,D-1$, we can efficiently estimate each $\mu_j$ and each $C^{\prime}_{jk}$ in time independent of $M$, which would ruin the exponential speedup potential of the proposed qGAD algorithm over the classical GAD algorithm. However, as shown in the equations~\eqref{eq:qOracleSgn} and \eqref{eq:qOracleAbs}, the proposed algorithm only assume quantum access to each entry of all data points $\mathbf{x}_i$, neither their quantum states $\ket{\mathbf{x}_{.j}}$  nor the norms $\norm{\mathbf{x}_{\cdot j}}$. This means it is reasonable to assume the operation (\romannumeral 1), while the operations (\romannumeral 2) and (\romannumeral 3) do not apply. Without operations (\romannumeral 2) and (\romannumeral 3), quantum-inspired classical sampling technique cannot efficiently estimate the inner product of any two vectors \cite{Tang19,Tang23}. Therefore, SQ data access model is not an appropriate analogy of the input data model in our qGAD algorithm, and thus our algorithm is resilient against quantum-inspired classical sampling.

\begin{table}[!t]
    \footnotesize
    \caption{Comparison among the classical GAD, prior quantum algorithms for GAD, and the proposed quantum algorithm for GAD. Here qGAD stands for quantum GAD, the symbol "/" means inexistent or not explicitly given.}
    \label{tab1}
    \begin{tabular*}{\textwidth}{cccccc}
    \toprule
       Algorithms & Data type &Mean centered & Features correlated & Circuit depth and size & Time complexity \\\hline
      Classical GAD  & Classical &Not required &Not required  &/ & $\mathcal{O}(D^2(M+D))$\\
      qGAD\cite{LSL19} & Quantum &Required & Not required &Exponential & / \\
      qGAD\cite{GLL22} & Classical &Not required & Required &Exponential & $\mathcal{O}\left(\frac{\log(MD)D^3}{\epsilon^3}\right)$\\
      The proposed qGAD &Classical & Not required & Not required &Linear & $\mathcal{O}\left(\frac{D^2\kappa^2(D+\kappa)^2\log(M)}{\epsilon^2}\right)$\\
    \bottomrule
    \end{tabular*}
    \end{table}

\subsection{Data practicality}
 The prior qGAD algorithms \cite{LSL19,GLL22} suffer from at least one of the three restrictions for input data: (a) the vector of each data point is required to be normalized and presented in a quantum state (i.e., quantum data) \cite{LSL19}; (b) the data is required to be mean centered which means each $\mu_j=0$ \cite{LSL19}; (c) each data feature is required to be uncorrelated with others which implies the covariance matrix $C$ is diagonal \cite{GLL22}. In contrast, our algorithm relaxes these requirements and handle generic input data; see Table \ref{tab1} for details. In this sense, the proposed qGAD algorithm is more practical.

\section{Conclusions}\label{sec:6}
We have proposed a qGAD algorithm to implement the central task of GAD, namely attaining the mean value vector $\bm{\mu}$ and the covariance matrix $C$ that determines the probability density function $p(\mathbf{x})$ of the hidden multivariate Gaussian distribution. In this algorithm, AFQSP is adopted to estimate the magnitude of every element of $\bm{\mu}$ and $C$, and the technique of Hadamard sign test is developed to further retrieve their signs. The proposed qGAD algorithm is exponentially faster than the classical counterpart when addressing the low-dimensional datasets with well-conditioned data matrices. Compared with the prior qGAD algorithms building on amplitude estimation, the quantum circuit of our algorithm has significantly shallower depth and thus is more favorable on the near-term quantum devices. In addition, our algorithm works for more generic input data, without requiring the data to be mean centered or feature correlated as required in the prior qGAD algorithms. The proposed qGAD algorithm exemplifies and may inspire more applications of AFQSP to narrowing the gap between quantum machine learning and reality.

\Acknowledgements{This work is supported by National Natural Science
Foundation of China (Grant Nos. 62006105,62301454, 62272206), China Postdoctoral Science Foundation (Grant No. 2023M731429),  Jiangxi Provincial Natural Science Foundation (Grant No. 20202BABL212004), Fundamental Research Funds for the Central Universities (Grant No. SWU-KQ22049), and the Natural Science Foundation of Chongqing (Grant No. CSTB2023NSCQ-MSX0739)}



\begin{appendix}
\section{Proof of Theorem 1}
\label{App:1}
Given a normalized single-qubit state $\ket{\varphi}=\alpha \ket{1}+\beta \ket{0}$ with real amplitudes $\alpha$ and $\beta$ and with $\alpha^2+\beta^2=1$,  one can implement the ``Hadamard+measure" procedure to determine the sign of $\alpha\beta$, that is, to determine whether $\alpha\beta\geq 0$ or $\alpha\beta<0$. Specifically, one can perform a Hadamard gate on this qubit, and then measure it $N_s$ times to see the outcome $\ket{1}$ with success probability $P_s:=\frac{1}{2}-\alpha\beta$. Letting $\hat{P}_s$ denote the actual proportion of getting the outcome $\ket{1}$, if $\hat{P}_s \leq 1/2$, $\alpha\beta\geq 0$ holds; otherwise, we believe  $\alpha\beta < 0$.

Now we analyze the scaling of $N_s$ when we want to correctly acquire the sign of $\alpha\beta$ with a probability greater than or equal to $1-\delta$. As we know, $N_s\hat{P}_s$ is a random variable in the binomial distribution $B(N_s,P_s)$, so $\hat{P}_s$ has expected value $\mathbb{E}(\hat{P}_s)=\frac{1}{2}-\alpha\beta$ and variance $\frac{\frac{1}{4}-\alpha^2\beta^2}{N_s}$. Consider two cases: $\alpha\beta\geq 0$ and $\alpha\beta < 0$. For the former case with $\alpha\beta\geq 0$, the probability of getting the right sign of $\alpha\beta$ is 
\begin{align}
\mathrm{Prob}\left(\hat{P}_s\leq 1/2\right)&= \mathrm{Prob}\left(\hat{P}_s-\mathbb{E}(\hat{P}_s) \leq \alpha\beta\right)\\
&=1- \mathrm{Prob}\left(\hat{P}_s-\mathbb{E}(\hat{P}_s) > \alpha\beta\right)\\
&\geq 1-\frac{\frac{1}{4}-\alpha^2\beta^2}{\frac{1}{4}-\alpha^2\beta^2+N_s\alpha^2\beta^2},
\label{ieq:a3}
\end{align} 
due to Cantelli inequality. So, if we require the probability to be greater than or equal to $1-\delta$, $N_s$ should satisfy  
\begin{align}
N_s \ge \frac{(1-\delta)(1-4\alpha^2\beta^2)}{4\delta\alpha^2\beta^2}.
\label{ieq:a4}
\end{align} 
Thus, $N_s=\mathcal{O}\left(\frac{1}{\delta}\left(\frac{1}{\alpha^2}+\frac{1}{\beta^2}\right)\right)$ is sufficient. 

For the other case $\alpha\beta< 0$, the sign of $\alpha\beta$ is acquired correctly with probability
\begin{align}
\mathrm{Prob}\left(\hat{P}_s > 1/2\right)&= \mathrm{Prob}\left(\hat{P}_s-\mathbb{E}(\hat{P}_s) > \alpha\beta\right)\\
&= 1- \mathrm{Prob}\left(\hat{P}_s-\mathbb{E}(\hat{P}_s) \leq \alpha\beta\right),
\end{align} 
which would give rise to the same inequalities as \eqref{ieq:a3} and \eqref{ieq:a4} according to Cantelli inequality. 

Putting the above two cases together,  the sign of $\alpha\beta$ can be determined with probability $\ge 1-\delta$ using $\mathcal{O}\left(\frac{1}{\delta}\left(\frac{1}{\alpha^2}+\frac{1}{\beta^2}\right)\right)$ copies of $\ket{\varphi}$.

\end{appendix}


\begin{thebibliography}{99}
  \bibitem{Shor94} P. W. Shor, Algorithms for quantum computation: Discrete logarithms and factoring, in \emph{Proceedings of the 35th Annual Symposium on the Foundations of Computer Science}, edited by S. Goldwasser (IEEE, Los Alamitos, California, 1994), pp.124--134.
  \bibitem{Grover97} L. K. Grover, Quantum mechanics helps in searching for a needle in a haystack, Phys. Rev. Lett. \textbf{79}, 325 (1997).
  \bibitem{HHL09} A. W. Harrow, A. Hassidim, and S. Lloyd, Quantum algorithm for linear systems of equations, Phys. Rev. Lett. \textbf{103}, 150502 (2009).
  \bibitem{QML17} J. Biamonte, P. Wittek, N. Pancotti, P. Rebentrost, N. Wiebe, and S. Lloyd, Quantum machine learning, Nature \textbf{549}, 195–202 (2017).  
  \bibitem{DWGetal25} Y. Du, X. Wang, N. Guo, Z. Yu, Y. Qian, K. Zhang, and D. Tao, Quantum machine learning: A hands-on tutorial for machine learning practitioners and researchers, arXiv:2502.01146 (2025).
  \bibitem{CK24} G. Cho, and D. Kim, Machine learning on quantum experimental data toward solving quantum many-body problems, Nature Communications \textbf{15}, 7552 (2024).

  \bibitem{LMR13} S. Lloyd, M. Mohseni, P. Rebentrost, Quantum algorithms for supervised and unsupervised machine learning, arXiv:1307.0411, 2013.
  \bibitem{PL13} K. L. Pudenz and D. A. Lidar, Quantum adiabatic machine learning, Quantum Inf. Process. \text12, 2027 (2013).
  \bibitem{QSVM14} P. Rebentrost, M. Mohseni, S. Lloyd, Quantum support vector machine for big data classification, Phys. Rev. Lett. \textbf{113}, 130503 (2014).
  \bibitem{CD16} I. Cong, L. Duan, Quantum discriminant analysis for dimensionality reduction and classification, New J. Phys. \textbf{18}, 073011 (2016).
  \bibitem{SFP17} M. Schuld, M. Fingerhuth and F. Petruccione, Implementing a distance-based classifier with a q uantum interference circuit, Europhysics Letters \textbf{119}, 60002 (2017).
  \bibitem{SP18} Schuld and F. Petruccione, Quantum ensembles of quantum classifiers, Sc. Rep. \textbf{8}, 2772 (2018).
  \bibitem{SK19} M. Schuld and N. Killoran, Quantum machine learning in feature Hilbert spaces, Phys. Rev. Lett. 122, 040504 (2019).
  \bibitem{HCTetal19}V. Havl\'{i}\^{c}ek, A. D. C\'{o}rcoles, K. Temme1, A. W. Harrow, A. Kandala, J. M. Chow, and Jay M. Gambetta, Supervised learning with quantum-enhanced feature spaces, Nature \textbf{567}, 209 (2019).
  \bibitem{SLWQWG24} Y. Song, J. Li, Y. Wu, S. Qin, Q. Wen, and F. Gao, A resource-efficient quantum convolutional neural network, Front Phys \textbf{12}, 1362690 (2024).

  \bibitem{ABG13} E. Aimeur, G. Brassard, S. Gambs, Quantum speed-up for unsupervised learning, Mach. Learn. 90, 261 (2013).
  \bibitem{KLLP19} I. Kerenidis, J. Landman, A. Luongo, and A. Prakash, q-means: A quantum algorithm for unsupervised machine learning, Advances in neural information processing systems, 32 (2019).
  \bibitem{AP21} Davis Arthur and Prasanna Date, Balanced k-means clustering on an adiabatic quantum computer,
  Quantum Information Processing \textbf{20}, 294 (2021).
  \bibitem{WSZ22} Z. Wu, T. Song, and Y. Zhang, Quantum k-means algorithm based on Manhattan distance, Quantum Information Processing \textbf{21}, 19 (2022).

  \bibitem{WBL12} N.Wiebe, D. Braun, S. Lloyd, Quantum algorithm for data fitting, Phys. Rev. Lett. \textbf{109}, 050505 (2012).
  \bibitem{SSP16}M. Schuld, I. Sinayskiy, F. Petruccione, Prediction by linear regression on a quantum computer, Phys. Rev. A \textbf{94}, 022342 (2016).
  \bibitem{Wang17} G. Wang, Quantum algorithm for linear regression, Phys. Rev. A 96, 012335 (2017).
  \bibitem{LZ17}Y. Liu and S. Zhang, Fast quantum algorithms for least squares regression and statistic leverage scores, Theor. Comput. Sci. \textbf{657}, 38 (2017).
  \bibitem{YGLetal19}C.-H. Yu, F. Gao, C. Liu, D. Huynh, M. Reynolds, and J. Wang, Quantum algorithm for visual tracking, Phys. Rev. A \textbf{99}, 022301 (2019).
  \bibitem{YGW21} C.-H. Yu, F. Gao, and Q.-Y. Wen. An improved quantum algorithm for ridge regression, IEEE Trans. Knowl. Data Eng. \textbf{33}, 858 (2021).
  \bibitem{CMP23} S. Chakraborty, A. Morolia, A. Peduri, A, Quantum regularized least squares, Quantum \textbf{7}, 988 (2023).
  \bibitem{QARM16} C.-H. Yu, F. Gao, Q.-L. Wang, and Q.-Y. Wen, Quantum algorithm for association rules mining, Phys. Rev. A \textbf{94}, 042311 (2016).
  \bibitem{Y22} C.-H. Yu. Experimental implementation of quantum algorithm for association rules mining. IEEE J. Emerging. Sel. Top Circuits Syst. \textbf{12}, 676-684 (2022).

  \bibitem{QPCA14} S. Lloyd, M. Mohseni, P. Rebentrost, Quantum principal component analysis, Nat. Phys. \textbf{10}, 631 (2014).
  \bibitem{DYXL19} B. Duan, J. Yuan, J. Xu, and D. Li, Quantum algorithm and quantum circuit for A-optimal projection: Dimensionality reduction, Phys. Rev. A \textbf{99}, 032311 (2019).
  \bibitem{YGLW19} C.-H. Yu, F. Gao, S. Lin, and J. Wang, Quantum data compression by principal component analysis, Quantum Information Processing \textbf{18}, 249 (2019).
  \bibitem{SDPC21} A. Sornsaeng, N. Dangniam, P. Palittapongarnpim, and T. Chotibut, Quantum diffusion map for nonlinear dimensionality reduction, Phys. Rev. A \textbf{104}, 052410 (2021).
  \bibitem{FGLX24} W. Feng, G. Guo, S. Lin,  and Y. Xu, Quantum Isomap algorithm for manifold learning, Phys. Rev. Appl. \textbf{22}, 014049 (2024).

  \bibitem{CBK09} V. Chandola, A. Banerjee, and V. Kumar, Anomaly detection: A survey, ACM Comput. Surv. \textbf{41}, 1 (2009).
  \bibitem{HGY22} W. Hilal, S. A. Gadsden, and J. Yawney, Financial fraud: a review of anomaly detection techniques and recent advances, Expert Syst. Appl. \textbf{193}, 116429 (2022).
  \bibitem{ASW21} Z. Ahmad, A. Shahid Khan, C. Wai Shiang,  and F. Ahmad, Network intrusion detection system: A systematic study of machine learning and deep learning approaches, Trans. Emerging Telecommun. Technol. \textbf{32}, 4150 (2021).
  \bibitem{FGD21} T. Fernando, H. Gammulle, S. Denman, S. Sridharan, and C. Fookes, Deep learning for medical anomaly detection–a survey, ACM Comput. Surv. \textbf{54}, 1 (2021).
  \bibitem{DM16} P.-N. Tan, M. Steinbach, and V. Kumar, \emph{Introduction to data mining} (Pearson Education India, 2016).

  \bibitem{LR18} N. Liu and P. Rebentrost, Quantum machine learning for quantum anomaly detection, Phys. Rev. A \textbf{97}, 042315 (2018).
  \bibitem{GLP23} M. C. Guo, H. L. Liu, S. J. Pan, W. M. Li, F. Gao, X. Y. Huang, and Q. Y. Wen, Quantum Algorithm for Anomaly Detection of Sequences, Adv. Quantum Technol. \textbf{6}, 2300082 (2023).
  \bibitem{GPL23} M. Guo, S. Pan, W. Li, F. Gao, S. Qin, X. Yu, and Q. Wen, Quantum algorithm for unsupervised anomaly detection, Physica A \textbf{625}, 129018 (2023).
  \bibitem{RYW25} H. M. Rao, C. H. Yu, Y. P. Wu, D. X. Liu, and X. P. Liu, Quantum algorithm for angle-based anomaly detection (in Chinese), Sci. Sin. Phys. Mech. Astron. \textbf{55}, 240307 (2025).
  \bibitem{CMD24} S. Corli, L. Moro, D. Dragoni, M. Dispenza, and E. Prati, Quantum machine learning algorithms for anomaly detection: A review, Future Gener. Comput. Syst, 107632 (2024).

  \bibitem{LSL19} J. M. Liang, S. Q. Shen, M. Li, and L. Li, Quantum anomaly detection with density estimation and multivariate Gaussian distribution, Phys. Rev. A \textbf{99}, 052310 (2019).
  \bibitem{GLL22} M. Guo, H. Liu, Y. Li, W. Li, F. Gao, S. Qin and Q. Wen, Quantum algorithms for anomaly detection using amplitude estimation, Physica A \textbf{604}, 127936 (2022).

  \bibitem{NISQ18} J. Preskill, Quantum computing in the NISQ era and beyond, Quantum \textbf{2}, 79 (2018).
  
  \bibitem{QST19} Y. R. Sanders, G. H. Low, A. Scherer, and D. W. Berry, Black-Box Quantum State Preparation without Arithmetic, Phys. Rev. Lett. \textbf{122}, 020502 (2019).

  \bibitem{RC04} S. A. Cuccaro, T. G. Draper, S. A. Kutin, and D. P. Moulton, A new quantum ripple-carry addition circuit, arXiv:quant-ph/0410184 (2004).
  \bibitem{Draper00} T. G. Draper, Addition on a Quantum Computer, arXiv:quant-ph/0008033 (2000).
  \bibitem{RG17} L. Ruiz-Perez , and J. C. Garcia-Escartin, Quantum arithmetic with the quantum Fourier transform, Quantum Inf. Process. \textbf{16}, 152 (2017).
  \bibitem{Sahin20} E. \c{S}ahin, Quantum arithmetic operations based on quantum Fourier transform on signed integers, Int. J. Quantum Inf. \textbf{18}, 2050035 (2020).
  \bibitem{Ferraz22} F C. Ferraz, Quantum algorithm based on quantum Fourier transform for register-by-constant addition, arXiv:2207.05309 (2022).

  \bibitem{Grover00} L. K. Grover, Synthesis of Quantum Superpositions by Quantum Computation, Phys. Rev. Lett. \textbf{85}, 1334 (2000).

  \bibitem{QRAM08} V. Giovannetti, S. Lloyd, and L. Maccone, Quantum random access memory, Phys. Rev. Lett. \textbf{100}, 160501 (2008).
  \bibitem{GLM08} V. Giovannetti, S. Lloyd, and L. Maccone, Architectures for a quantum random access memory, Phys. Rev. A \textbf{78}, 052310 (2008).
  \bibitem{HLGJ21} C. T. Hann, G.  Lee, S. M. Girvin, L. Jiang, Resilience of quantum random access memory to generic noise, Prx Quantum, \textbf{2}, 020311 (2021).
  
  \bibitem{IR08} I. C. F. Ipsen, and R. Rehman, Perturbation bounds for determinants and characteristic polynomials. SIAM Journal on Matrix Analysis and Applications, \textbf{30}, 762-776 (2008)
  \bibitem{Higham02} N. J. Higham, \emph{Accuracy and Stability of Numerical Algorithms: Second Edition} (Society for Industrial and Applied Mathematics, 2002)
  
  \bibitem{Tang19} Tang E. A quantum-inspired classical algorithm for recommendation systems. In: Proceedings of the 51st Annual ACM SIGACT Symposium on Theory of Computing. Phoenix, 2019. 217–228
  \bibitem{Tang21} Tang E., Quantum principal component analysis only achieves an exponential speedup because of its state preparation assumptions, Phys Rev Lett, 2021, 127: 060503
  \bibitem{GST22} A Gily\'{e}n , Z. Song, E. Tang, An improved quantum-inspired algorithm for linear regression. Quantum, 2022, 6: 754
  \bibitem{Tang22} Tang E. Dequantizing algorithms to understand quantum advantage in machine learning. Nat Rev Phys, 2022, 4: 692–693
  \bibitem{Tang23} Tang, E. Quantum machine learning without any quantum. University of Washington, 2023.
  
  \bibitem{SV24}A. Shukla, P. Vedula, An efficient quantum algorithm for preparation of uniform quantum superposition states, Quantum Inf. Process. \textbf{23}, 38 (2024). 

\end{thebibliography}
\end{document}